\documentclass[a4paper,twocolumn,11pt]{quantumarticle}
\pdfoutput=1
\usepackage[utf8]{inputenc}
\usepackage[english]{babel}
\usepackage[T1]{fontenc}
\usepackage{amsmath}
\usepackage{hyperref}

\usepackage{amsmath,amssymb,amsthm,mathtools}
\usepackage{graphicx}
\usepackage{bm}
\usepackage{xcolor}
\usepackage{physics}
\usepackage{booktabs}

\usepackage{tikz}
\usepackage{lipsum}

\newtheorem{theorem}{Theorem}[section]

\newcommand{\IB}{I_{\mathrm{B}}(A;B|C)}
\newcommand{\QCMI}{I_{\mathrm{B}}(A;B|C)}
\newcommand{\QCMIsym}{I(A:B|C)}
\newcommand{\Ic}{I_{\mathrm{c}}}
\newcommand{\vN}{S}
\newcommand{\rhoABC}{\rho_{ABC}}

\newcommand{\rhoBCx}{\rho^{x}_{BC}}
\newcommand{\phizero}{\varphi_0}
\newcommand{\Nzero}{N_0}
\newcommand{\NzeroA}{N_0^A}
\newcommand{\NzeroC}{N_0^C}
\newcommand{\Tc}{T_{\mathrm{c}}}
\newcommand{\cs}{c_{\mathrm{s}}}
\newcommand{\vLR}{\tilde{v}_{\mathrm{LR}}}
\newcommand{\veff}{v_{\mathrm{eff}}}
\newcommand{\ord}{\mathcal{O}}

\begin{document}

\title{Bosonic Quantum Conditional Mutual Information and Bose-Einstein Condensation}

\author{Anupam Ghosh}
\email{ghosh@cs.cas.cz}
\affiliation{Department of Complex Systems, Institute of Computer Science of the Czech Academy of Sciences, Prague 18200, Czech Republic}
\orcid{0000-0002-7778-7572}

\maketitle

\begin{abstract}
  This article aims to introduce the bosonic asymmetric quantum conditional mutual information (QCMI), $\IB$, as a rigorously defined information-theoretic quantity for continuous-variable (bosonic) systems. After defining $\IB$ through quantum instruments on infinite-dimensional Fock spaces, we prove that Bose-Einstein condensation (BEC) induces a logarithmic divergence $\IB\!\geq\!\alpha\log_2(\NzeroC\!+\!1)$, which further establishes $\IB$ as a quantum-information order parameter for BEC with power-law scaling near the critical point. We provide the complete five-step proof of this result, connecting measurement back-action, off-diagonal long-range order, and Jensen's inequality. In the condensed phase, the effective causal propagation velocity is set by the Bogoliubov sound speed $c_s$ in the long-wavelength limit, and vanishes continuously at the superfluid--Mott-insulator transition. Applications to composite bosons, the BEC--BCS crossover, and a quantum-gas-microscope measurement protocol are presented. Finally, we demonstrate how the reduced bosonic QCMI serves as a finite-size scaling probe analogous to the Binder cumulant. 
\end{abstract}

\section{Introduction}
\label{sec:intro}
Causal inference---identifying \emph{directional} information flow between interacting systems---is a critical topic of modern physics and data science~\cite{pearl09,oreshkov12,leifer13}. In classical dynamics, a specific form of conditional mutual information (CMI) $\Ic(A;B|C)$~\cite{palus01,palus07}, among many~\cite{granger69,schreiber00,liang16}, serves as an asymmetric causal index. Hence, $\Ic(A;B|C)$ distinguishes the direction $A\to B$ from $B\to A$. A recent Letter~\cite{ghosh25} established a rigorous quantum generalization---the \emph{asymmetric} QCMI---by conditioning on post-measurement states generated by a quantum instrument acting on subsystem $A$. The asymmetric QCMI distinguishes the directions $A\to B$ and $B\to A$ by quantifying how much the measurement of $A$ reshapes the correlations between $B$ and $C$. Applied to spin-$\tfrac{1}{2}$ chains, the QCMI propagates at a finite effective velocity bounded by the Lieb-Robinson (LR) speed~\cite{lieb72}, with coherent oscillations reflecting the Jordan-Wigner dispersion~\cite{jordan28,sur20}.
An important class of physical systems---ultracold atomic gases~\cite{anderson95,davis95,bradley95}, optical lattices~\cite{jaksch98,greiner02}, and photonic condensates~\cite{klaers10}---consists of bosons. Bosonic systems differ from spin chains in two ways that are relevant to causal inference: First, the local Hilbert spaces are \emph{infinite-dimensional} Fock spaces, which requires a careful generalization of quantum instruments beyond the qubit projectors~\cite{ghosh25}. Second, and more fundamentally, bosonic ground states can spontaneously break $\mathrm{U}(1)$ symmetry via BEC, producing off-diagonal long-range order (ODLRO)~\cite{penrose56} that coherently links arbitrarily distant subsystems. This macroscopic quantum coherence has no analogue in spin chains and, as we show, significantly amplifies inter-subsystem causal influence. The extension to bosonic systems is also relevant as quantum-gas microscopes now achieve single-site-resolved Fock-state detection in Hubbard-regime lattices~\cite{bakr09,sherson10}, making the bosonic QCMI experimentally accessible.
This paper makes three main contributions: First, we define the bosonic asymmetric QCMI $\QCMI$ through quantum instruments on symmetrized Fock spaces and establish its key properties. Second, we prove that BEC produces a logarithmic divergence in $\QCMI$ proportional to $\log \Nzero$, making $\QCMI$ a quantum-information order parameter for the phase transition. Third, we show that the effective causal propagation velocity in the condensed phase is controlled by the Bogoliubov sound speed $\cs$, which vanishes at the SF-MI transition of the Bose-Hubbard model~\cite{fisher89}, providing a dynamical signature of the quantum phase transition.
The structure of this paper is as follows: Section~\ref{sec:model} presents the definition of bosonic QCMI and examines its operational properties. Section~\ref{sec:gaussian} develops the Gaussian covariance-matrix formalism, including the closed-form expression for homodyne detection and the infrared divergence at BEC. Section~\ref{sec:theorem} states and outlines the proof of the BEC Enhancement Theorem. Section~\ref{sec:order} investigates the role of QCMI as an order parameter. Section~\ref{sec:causal} addresses causal propagation through the Bogoliubov spectrum and the LR bound. Section~\ref{sec:applications} describes relevant applications. Section~\ref{sec:discussion} considers implications and outlines potential future research directions. Finally, Sec.~\ref{sec:conclusions} provides concluding remarks.

\section{Bosonic QCMI: Definition and Properties}
\label{sec:model}
Consider $N$ identical bosons on an $L$-site lattice with creation and annihilation operators $a_i^\dagger$, $a_i$ satisfying $[a_i,a_j^\dagger]=\delta_{ij}$. The local Hilbert space is the Fock space $\mathcal{F}_i=\mathrm{span}\{|n\rangle_i:n\geq 0\}$, and the full space is $\mathcal{H}=\mathrm{Sym}\!\bigl(\bigotimes_{i=1}^L\mathcal{F}_i\bigr)$. We partition the $L$ sites into three disjoint regions $A$, $B$, $C$ and denote the tripartite density operator $\rhoABC$.

The symmetric QCMI,
\begin{equation}
	\QCMIsym=\vN(\rho_{AC})+\vN(\rho_{BC})-\vN(\rho_C)-\vN(\rhoABC),
	\label{eq:sym}
\end{equation}
with $\vN(\cdot)$ the von Neumann entropy, satisfies $\QCMIsym=I(B:A|C)$ and therefore encodes \emph{no} causal direction.
\subsection{Definition}
Asymmetry is imposed by a quantum instrument $\mathcal{M}_A=\{M_x\}_{x\in\mathcal{X}}$ on $A$, with
$\sum_x M_x^\dagger M_x=\mathbb{I}_A$. The post-measurement state on $BC$ for outcome $x$ is
\begin{equation}
	\rhoBCx=\frac{\mathrm{Tr}_A\!\bigl[
		(M_x\otimes\mathbb{I}_{BC})\rhoABC
		(M_x^\dagger\otimes\mathbb{I}_{BC})\bigr]}
	{p(x)},
	\label{eq:postmeas}
\end{equation}
with $p(x)=\mathrm{Tr}[(M_x^\dagger M_x\otimes\mathbb{I}_{BC}) \rhoABC]$. The bosonic asymmetric QCMI is then
\begin{equation}
	\QCMI:=\int_{\mathcal{X}}p(x)\, I(B:C)_{\rhoBCx}\,\mathrm{d}\mu(x),
	\label{eq:bqcmi}
\end{equation}
where $I(B:C)_{\rhoBCx}=\vN(\rho^x_B)+\vN(\rho^x_C)-\vN(\rhoBCx)$ is the quantum mutual information of the post-measurement state, and $\mathrm{d}\mu$ is counting (discrete) or Lebesgue (continuous) measure. Equation~\eqref{eq:bqcmi} generalizes Eq.~(7) of Ref.~\cite{ghosh25} to infinite-dimensional Fock spaces. Intriguingly, the measurement on $A$ is the \emph{only} mechanism that can create new $BC$-correlations. Hence, a positive $\IB$ value implies the causal influence: $A\to B$.
Three different classes are physically natural: (a) number-resolving projectors $M_n=|n_A\rangle\langle n_A|$
(Fock-space analogue of the qubit projectors of Ref.~\cite{ghosh25}); (b) homodyne measurements $M_\alpha=|\alpha\rangle\langle\alpha|/\pi$ via coherent-state POVMs, suited to Gaussian states; and (c) particle-number-conserving instruments, appropriate when global U(1) symmetry must be preserved.

\begin{table*}
	\caption{\label{tab:notation} Principal symbols used throughout this paper.}
	\centering
	\begin{tabular}{ll}
		\hline\hline
		\textrm{Symbol} & \textrm{Meaning} \\
		\hline
		$I_B(A;B|C)$ & Bosonic asymmetric QCMI [Eq.~\eqref{eq:bqcmi}] \\
		$\mathcal{M}_A=\{M_x\}$ & Quantum instrument on region $A$ \\
		$\rho^x_{BC}$ & Post-measurement state on $BC$ after outcome $x$ \\
		$g(\nu)$ & Symplectic entropy function [Eq.~\eqref{eq:gaussent}] \\
		$\nu_k\geq 1$ & Symplectic eigenvalues of the covariance matrix \\
		$\sigma$ & Covariance matrix (CM) \\
		$f_0(T)$ & Global condensate fraction $N_0/N$ \\
		$w_X$ & Condensate weight in region $X$: $\int_X|\varphi_0|^2\,dV$ \\
		$N_0^C$ & Expected condensate particle count in region $C$ \\
		$\alpha$ & BEC prefactor $g(w_B,w_C)\cdot w_C/(w_B+w_C)$ [Eq.~\eqref{eq:alpha}] \\
		$q_B(T)$ & Reduced bosonic QCMI order parameter [Eq.~\eqref{eq:orderparam}] \\
		$c_s$ & Bogoliubov sound speed $\sqrt{gn_0/m}$ \\
		$\xi$ & Healing length $\hbar/\sqrt{2mgn_0}$ \\
		$v_g(k)$ & Bogoliubov group velocity [Eq.~\eqref{eq:vg}] \\
		$\tilde{v}_{\mathrm{LR}}$ & Bosonic Lieb--Robinson velocity $2eJn_{\max}$ [Eq.~\eqref{eq:lr-boson}] \\
		$\xi_{\mathrm{MI}}$ & MI correlation length \\
		$\nu_{\mathrm{MF}}$ & Mean-field dynamical exponent [Eq.~\eqref{eq:cs-vanish}] \\
		\hline\hline
	\end{tabular}
\end{table*}

\subsection{Properties}
The bosonic QCMI satisfies five key properties. Full proofs are given in the Supplementary Material (SM).
\begin{enumerate}
	\item \textit{Non-negativity.} $\QCMI \geq 0$ for every bosonic state $\rhoABC$ and every quantum instrument $\mathcal{M}_A$. This follows from the strong subadditivity (SSA) of the von Neumann entropy~\cite{lieb73}, which implies $I(B:C)_{\rhoBCx} \geq 0$ for every outcome $x$.
	\item \textit{Asymmetry.} $I_{\rm B}(A;B|C) \neq I_{\rm B}(B;A|C)$ in general. The signed difference $\Delta I_{\rm B} = I_{\rm B}(A;B|C) - I_{\rm B}(B;A|C)$ is the causal asymmetry index; $\Delta I_{\rm B} > 0$ indicates net causal dominance $A\to B$. Asymmetry arises because the measurement back-action on $A$ is physically different from back-action on $B$.
	\item \textit{Classical reduction.}	When $\rhoABC$ is diagonal in the Fock basis and $\mathcal{M}_A$ consists of number-resolving projectors, $\QCMI$ reduces to the classical CMI~\cite{palus01}.
	\item \textit{Hard-core limit.} With at most one boson per site ($n_{\max}=1$), $\QCMI$ reduces via the Jordan-Wigner (JW) transformation~\cite{jordan28} to the QCMI of Ref.~\cite{ghosh25} for spin-$\tfrac{1}{2}$ chains. This is the key consistency check that connects the bosonic generalization back to the original framework.
	\item \textit{Data-processing inequality.} For any quantum channel $\Phi_B$ on $B$, $I_{\rm B}(A;\Phi_B(B)|C) \leq \QCMI$. This ensures that locally processing $B$'s data cannot create new causal influence from $A$.
\end{enumerate}
For convenience, Table~\ref{tab:notation} lists the principal symbols used throughout this paper.
%

\section{Gaussian States and Analytical Results}
\label{sec:gaussian}
Gaussian states are the thermal and ground states of free-boson Hamiltonians, and they arise naturally in BEC mean-field theory, optical parametric processes, and Bogoliubov theory~\cite{holevo11}. Their key virtue is that the full quantum state is encoded in the covariance matrix (CM), a finite-dimensional object. 

A bosonic state $\rho$ on $L$ modes is Gaussian if its Wigner function is a Gaussian distribution on phase space. Equivalently, $\rho$ is fully characterized by the displacement vector $\langle \bm{r}\rangle$ and the CM~\cite{holevo11}
\begin{equation}
	\sigma_{ij}=\langle\{r_i,r_j\}\rangle-2\langle r_i\rangle\langle r_j\rangle,
\end{equation}
where $\bm{r}=(x_1,p_1,x_2,p_2,\ldots,x_L,p_L)^T$ are quadrature operators and $\{r_i,r_j\} = (r_i r_j + r_j r_i)$. The CM must satisfy the Heisenberg uncertainty relation
\begin{equation}
	\sigma + i\Omega \geq 0,
\end{equation}
where $\Omega = \bigoplus_{k=1}^L \bigl(\begin{smallmatrix}0&1\\-1&0\end{smallmatrix}\bigr)$ is the symplectic form. The resulting positive numbers $\{\nu_k \geq 1\}$ are called symplectic eigenvalues. By the Williamson decomposition~\cite{houde24}, every real, symmetric, positive-definite $\sigma$ can be brought to block-diagonal form $S \sigma S^T = \bigoplus_k \nu_k\, \mathbb{I}_2$ by a symplectic transformation $S\in\mathrm{Sp}(2L,\mathbb{R})$. The von Neumann entropy of a Gaussian state is determined by its symplectic spectrum~\cite{serafini04}:
\begin{equation}
	\vN(\rho)=\sum_k g(\nu_k),
	\label{eq:gaussent}
\end{equation}
where $g(\nu)=\tfrac{\nu+1}{2}\log_2\tfrac{\nu+1}{2}-\tfrac{\nu-1}{2}\log_2\tfrac{\nu-1}{2}$. For a single thermal mode with mean occupation $\bar{n}$, the CM is $\sigma = (2\bar{n}+1)\mathbb{I}_2$ and $\nu = (2\bar{n}+1)$. For a large $\nu$-value (i.e., $\bar{n}\to\infty$, the condensate mode in the condensed phase),
\begin{equation}
	g(\nu) \approx \log_2\nu,
\end{equation}
so modes with large occupation contribute logarithmically to the entropy.
\begin{figure}[htbp!]
	\centering
	\includegraphics[width=\columnwidth]{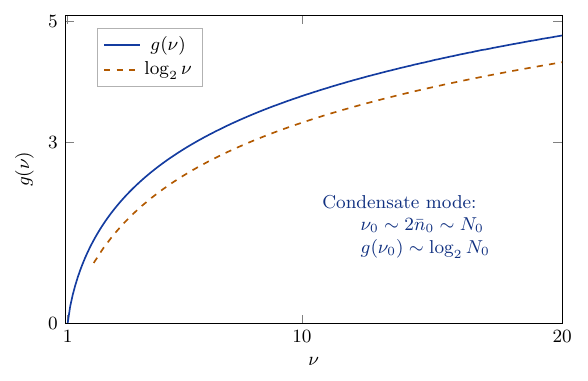}
	\caption{%
		\textbf{The Gaussian entropy function $g(\nu)$ and the condensate divergence.} The per-mode entropy $g(\nu)$ (blue) grows logarithmically for $\nu\gg 1$, with asymptote $\log_2\nu$ (orange dashed). For a pure state ($\nu=1$), $g=0$. In the condensed phase, the condensate mode has $\nu_0\approx (2\bar{n}_0 + 1) \sim N_0$, contributing $g(\nu_0)\sim\log_2 N_0$ to the entropy and driving the logarithmic divergence of $\IB$.}
	\label{fig:gfunc}
\end{figure}

Figure~\ref{fig:gfunc} displays the symplectic entropy function $g(\nu) = \frac{\nu+1}{2}\log_2\frac{\nu+1}{2} - \frac{\nu-1}{2}\log_2\frac{\nu-1}{2}$, which quantifies the von Neumann entropy of a single bosonic mode with symplectic eigenvalue $\nu \geq 1$. The function vanishes at $\nu = 1$, corresponding to a pure Gaussian state with zero entropy, and grows monotonically for $\nu > 1$ as the mode becomes increasingly mixed. For large $\nu$, the blue solid curve asymptotes to $\log_2 \nu$ (orange dashed line), recovering the classical entropy of a highly occupied, thermal-like mode. The physical significance of the BEC Enhancement Theorem (Theorem~\ref{thm:bec}) is indicated by the annotation: the condensate mode carries a symplectic eigenvalue $\nu_0 \sim 2\bar{n}_0 \sim N_0$, proportional to the macroscopic condensate occupation $N_0$, so that its symplectic entropy satisfies $g(\nu_0) \sim \log_2 N_0$. This logarithmic growth of $g(\nu_0)$ with $N_0$ that underlies the logarithmic divergence of the bosonic QCMI in the presence of BEC, i.e.,  conditioning the measurement on region $C$ accesses the condensate mode, and the entropy $g(\nu_0)$ sets the scale of the resulting back-action on $AB$.
\subsection{Post-measurement covariance matrix under homodyne detection}
For a tripartite Gaussian state $\rhoABC$ with CM partitioned in blocks, a homodyne measurement~\cite{eisert02} of the position quadrature $x_{A}$ on all sites of region $A$ yields the post-measurement CM on $BC$:
\begin{equation}
	\sigma^x_{BC}=\sigma_{BC} -\sigma_{A,BC}^T(\Pi_x\sigma_A\Pi_x)^{\rm MP}\sigma_{A,BC},
	\label{eq:postcm}
\end{equation}
where $\sigma_{A,BC} = (\sigma_{AB}, \sigma_{AC})$, $\Pi_x = \mathrm{diag}(\mathbb{I},\mathbf{0})$ projects onto the measured quadrature, and ${}^{\mathrm{MP}}$ denotes the Moore-Penrose pseudoinverse~\cite{eisert02}. The key fact is that $\sigma^x_{BC}$ is \emph{independent of the measurement outcome} $x$ for Gaussian states. This is the Schur-complement structure of the Gaussian conditional state. Therefore, the bosonic QCMI under homodyne measurement reduces to a function of the CM:
\begin{eqnarray}
	I_{\rm B}^{\rm hom}(A;B|C)&=&I(B:C)_{\sigma^x_{BC}}\nonumber\\
	&=& \sum_k g(\nu_k^B) + \sum_k g(\nu_k^C) - \sum_k g(\nu_k^{BC}),
	\label{eq:gaussIB}
\end{eqnarray}
where $\nu_k^X$ are the symplectic eigenvalues of the appropriate marginal. This outcome-independence is the Gaussian-state signature of BEC. When the condensate mode diverges ($\nu_0\to\infty$ as $T\to\Tc$), Eq.~\eqref{eq:gaussIB} diverges logarithmically by anticipating the Theorem~\ref{thm:bec} of Sec.~\ref{sec:theorem}.
\subsection{Infrared divergence of the covariance matrix at BEC}
For a free Bose gas on an $L$-site lattice with hopping Hamiltonian $H_0 = -J\sum_{\langle ij\rangle} (a_i^\dagger a_j + \mathrm{h.c.})$, the thermal CM element at inverse temperature $\beta = 1/T$ is
\begin{equation}
	\sigma^{xx}_{ij}
	= \frac{1}{L}\sum_k \cos[k(i-j)]
	\coth\!\frac{\beta\varepsilon_k}{2},
	\label{eq:freecm}
\end{equation}
where $\varepsilon_k = -2J\cos k$, and the sum runs over the $L$ quasi-momenta in the first Brillouin zone. In the normal phase ($T > \Tc$), all $\coth(\beta\varepsilon_k/2)$ are bounded and the CM is finite. In the condensed phase ($T < \Tc$), the $k=0$ mode has $\varepsilon_0 \to 0$ in the thermodynamic limit, giving
\begin{equation}
	\coth\!\frac{\beta\varepsilon_0}{2}
	\xrightarrow{\varepsilon_0\to 0}
	\frac{2T}{\varepsilon_0} \to \infty.
	\label{eq:ir-div}
\end{equation}
This \emph{infrared divergence} of the CM at $k=0$ is the Gaussian-state signature of BEC. The condensate mode acquires a symplectic eigenvalue 
\begin{equation}
	\nu_0 \approx 2\bar{n}_0 + 1 \sim \Nzero,
\end{equation}
which diverges with system size. Through Eq.~\eqref{eq:gaussIB}, this drives $g(\nu_0) \sim \log_2\Nzero$ and hence $\IB \to \infty$. This provides an independent, analytical derivation of the BEC divergence at the covariance matrix level and is consistent with the main theorem proved in Sec.~\ref{sec:theorem}. The mechanism is simple: the macroscopic occupation of the $k=0$ mode drives $\nu_0\to\infty$, which diverges through the symplectic entropy function $g(\nu)$ and feeds directly into $I_B^{\rm hom}(A;B|C)$ via Eq.~\eqref{eq:gaussIB}. Section~\ref{sec:theorem} makes this precise for general instruments.

\section{BEC Enhancement Theorem}
\label{sec:theorem}
BEC below $\Tc$ is characterized by macroscopic occupation of a single mode $\phizero$. The single-particle density matrix $\rho^{(1)}(r,r')=\langle a^\dagger(r)a(r')\rangle$ acquires eigenvalue $\Nzero=Nf_0(T)$, where $f_0(T)=[1-(T/\Tc)^{d/2}]_+$ is the condensate fraction in $d$ dimensions~\cite{huang08}; for the ideal 3D Bose gas this gives $f_0\propto(1-T/\Tc)^{3/2}$ near $\Tc$:
\begin{equation}
	\lim_{|r-r'|\to\infty}\rho^{(1)}(r,r')=\frac{\Nzero}{V}\neq 0
	\quad(T<\Tc).
	\label{eq:odlro}
\end{equation}
This off-diagonal long-range order (ODLRO) coherently couples arbitrarily distant regions $B$ and $C$ through the shared condensate mode. Ding and Yang~\cite{ding09} showed that ODLRO produces a $\sim\!\log\Nzero$ contribution to the \emph{symmetric} mutual information $I(A:B)$. Here, we extend this result to the \emph{asymmetric} bosonic QCMI, where measurement at $A$ \emph{exploits} the ODLRO to generate new $BC$-correlations that grow without bound as $\Nzero\to\infty$.

\begin{theorem}
	\label{thm:bec}
	Let $\rhoABC$ be the density operator of a free or weakly interacting Bose gas at temperature $T < \Tc$, with condensate occupation $\Nzero$ in mode $\phizero$. Let $\mathcal{M}_A$ be a number-resolving projective measurement on region $A$. Then:
	\begin{equation}
		\QCMI \geq \alpha\,\log_2\!\bigl(\NzeroC + 1\bigr) + \mathcal{O}(1),
		\label{eq:main}
	\end{equation}
	where $\NzeroC = \Nzero\!\int_C|\phizero(r)|^2\,\mathrm{d}V$ is the expected condensate count in the conditioning region $C$, and $\alpha > 0$ depends on the condensate wavefunction geometry. For $T > \Tc$ ($\Nzero = 0$), the divergence is absent and $\QCMI = \mathcal{O}(1)$.
\end{theorem}

Now, we prove the lower bound in the following five steps:

\emph{1. Decomposition of the density operator.}
In the condensed phase, the many-body density operator factorizes to leading order in $1/\Nzero$ as
\begin{equation}
	\rhoABC \simeq \rho_{\Nzero} \otimes \rho_{\mathrm{ex}},
	\label{eq:factorize}
\end{equation}
where $\rho_{\Nzero} = |\Nzero\rangle\langle\Nzero|_{\phizero}$ is the Fock state of $\Nzero$ particles in the condensate mode, and $\rho_{\mathrm{ex}}$ is the thermal state of excited particles with occupation $n_k = \ord(1)$. This factorization is exact for the ideal Bose gas and holds to $\ord(\Nzero^{-1/2})$ in the Bogoliubov approximation. The condensate mode decomposes across the three regions as 
\begin{equation}
	\phizero(r) = \sqrt{w_A}\,\psi_A(r) + \sqrt{w_B}\,\psi_B(r) + \sqrt{w_C}\,\psi_C(r),
\end{equation}
where $w_X = \int_X|\phizero(r)|^2\,\mathrm{d}V$ ($X\in\{A,B,C\}$, $w_A+w_B+w_C=1$) is the condensate weight in each region.

\emph{2. Outcome distribution.} A number-resolving measurement on region $A$ with outcome $n$ draws from the condensate sector with statistics $\mathrm{Binomial}(\Nzero, w_A)$. For $\Nzero\gg 1$ with $w_A$ fixed, this converges to a Poisson distribution with mean $\mu_A = \NzeroA$:
\begin{equation}
	p_n \approx e^{-\mu_A}\frac{\mu_A^n}{n!}, 
	\label{eq:poisson}
\end{equation}
where $\mu_A = \Nzero w_A = \NzeroA.$

\emph{3. Post-measurement ODLRO.} Given outcome $n$, the post-measurement state on $BC$ contains a surviving $(\Nzero - n)$-particle condensate in regions $B$ and $C$. The single-particle density matrix of this state retains ODLRO:
\begin{equation}
	\lim_{|r-r'|\to\infty}\rho^{(1),n}_{BC}(r,r') = \frac{\Nzero - n}{V_{BC}} \neq 0,
	\label{eq:ODLRO-post}
\end{equation}
with $(\Nzero - n) \gg 1$. The measurement back-action at $A$ reduces the condensate by $n$ particles but does not destroy ODLRO in $BC$. By the result of Ding and Yang~\cite{ding09}, ODLRO with condensate occupation $M$ generates a mutual information $\sim\log_2 M$ between any two regions sharing that mode:
\begin{equation}
	I(B:C)_{\rho^n_{BC}} \geq g(w_B, w_C)\,\log_2(\Nzero-n+1) + \ord(1),
	\label{eq:MI-lower}
\end{equation}
where $g(w_B, w_C) = H_{\mathrm{bin}}[w_B/(w_B+w_C)]$ is the binary entropy function.

\emph{4. Jensen's inequality.} Substituting the lower bound, Eq.~\eqref{eq:MI-lower}, into Eq.~\eqref{eq:bqcmi}:
\begin{equation}
	\IB \geq g(w_B,w_C) \sum_{n=0}^{\Nzero} p_n\,\log_2(\Nzero-n+1) + \ord(1).
\end{equation}
Since $\log_2$ is a concave function, Jensen's inequality gives
\begin{equation}
	\mathbb{E}_n[\log_2(\Nzero-n+1)] \geq \log_2\bigl(\mathbb{E}_n[\Nzero-n+1]\bigr).
	\label{eq:Jensen}
\end{equation}
Using the Poisson mean $\mathbb{E}_n[n] = \NzeroA$:
\begin{eqnarray}
	\mathbb{E}_n[\Nzero-n+1] = \Nzero(1-w_A)+1 \nonumber\\ \qquad = \Nzero(w_B+w_C)+1 \geq \NzeroC + 1.
	\label{eq:mean}
\end{eqnarray}

\emph{5. Above $\Tc$.} For $T > \Tc$, no single-particle mode is macroscopically occupied. All $n_k = \ord(1)$, so $I(B:C)_{\rho^n_{BC}} = \ord(1)$ for every $n$. Averaging over outcomes gives $\IB = \ord(1)$, and the logarithmic divergence is absent.

Combining Steps~1--4 and after defining 
\begin{equation}
	\alpha = g(w_B,w_C)\cdot \frac{w_C}{(w_B+w_C)},
	\label{eq:alpha}
\end{equation}
we obtain Eq.~\eqref{eq:main}, which completes the proof. $\square$
%

\begin{figure}[ht]
	\centering
	\includegraphics[width=\columnwidth]{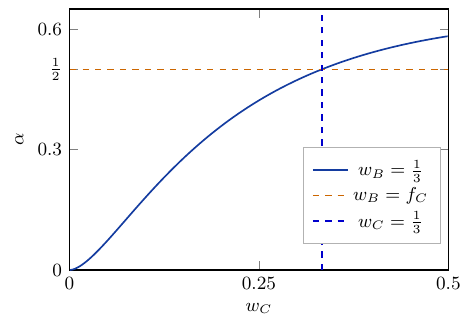}
	\caption{ \textbf{Geometry dependence of the BEC Enhancement prefactor $\alpha$.} The prefactor $\alpha$, as defined in Eq.~\eqref{eq:alpha}, governs the magnitude of the logarithmic divergence. The blue solid curve illustrates the dependence of $\alpha$ on $w_C$, with $w_B$ held fixed at $\tfrac{1}{3}$. The orange dashed line represents the symmetric case $w_B=w_C$, for which $\alpha = \tfrac{1}{2}$. The prefactor $\alpha$ attains its global maximum $\alpha_{\max}\approx 0.607$ at $w_C/(w_B+w_C)\approx 0.62$. As $w_C$ approaches zero, $\alpha$ tends to zero and the bound becomes trivial. The blue dashed line corresponds to $w_C = \tfrac{1}{3}$.}
	\label{fig:alpha}
\end{figure}
Note that the prefactor $\alpha$, Eq.~\eqref{eq:alpha}, has a geometric meaning. It is determined by two factors: First, $g(w_B,w_C) = H_{\mathrm{bin}}[w_B/(w_B+w_C)] \leq 1$ measures how evenly the condensate is split between $B$ and $C$—the mutual information between $B$ and $C$ is maximized when neither subsystem holds all the condensate. Second, $w_C/(w_B+w_C) \leq 1$ is the fraction of the $BC$-condensate that resides in the conditioning region $C$. The bound is tightest (largest $\alpha$) when $w_B = w_C$, giving $\alpha = \tfrac{1}{2}$, and vanishes when $w_C\to 0$. (Since there is no condensate in $C$, conditioning on $C$ provides no information about the condensate.) Finally, $\alpha$ exceeds $\tfrac{1}{2}$ for $w_C > w_B$. The variation of $\alpha$ at different combinations of $w_C$ and $w_B$ values is depicted in Fig.~\ref{fig:alpha}.
Finally, the main claim of this theorem is that BEC amplifies inter-subsystem causal influence without bound as the condensate grows. Intuitively, the condensate is a single quantum mode $\phizero$ that is delocalized across the entire system. When we measure $n$ particles in region $A$, the back-action does not stay local. Instead, it renormalizes the global condensate wavefunction, modifying the quantum state of regions $B$ and $C$ through their shared participation in $\phizero$. This measurement-induced update generates new correlations between $B$ and $C$ proportional to $(\Nzero-n)$, which grows with the condensate occupation $\Nzero$. Averaging over outcomes and applying Jensen's inequality converts this observation into a rigorous lower bound.
%
\section{QCMI as an Order Parameter for BEC}
\label{sec:order}
Theorem~\ref{thm:bec} motivates us to define the reduced bosonic QCMI per unit condensate:
\begin{equation}
	q_B(T)=\lim_{L\to\infty}\frac{\QCMI}{\log_2(\NzeroC+1)},
	\label{eq:orderparam}
\end{equation}
where $\NzeroC\propto L^{d/2}$ in the superfluid phase, so both numerator and denominator diverge and the ratio
remains finite. By construction, $q_B(T)>0$ if and only if $T<\Tc$, which implies a quantum-information signature of the BEC transition. Following Fig.~\ref{fig:orderparam}, $q_B$ tracks the condensate fraction, and near the critical point, i.e., when $T\to\Tc^-$,
\begin{equation}
	q_B(T)\propto f_0(T)\propto(1-T/\Tc)^\beta,
	\label{eq:scaling}
\end{equation}
with $\beta=3/2$ for the ideal 3D Bose gas, or the interacting exponent near the Berezinskii-Kosterlitz-Thouless
transition~\cite{berezinskii71,kosterlitz73}. This universal power-law scaling of causal information at the
BEC transition is a prediction that can be tested experimentally by measuring $\IB$ at several temperatures below $\Tc$ and fitting the exponent $\beta$.

\begin{figure}[ht]
	\centering
	\includegraphics[width=\columnwidth]{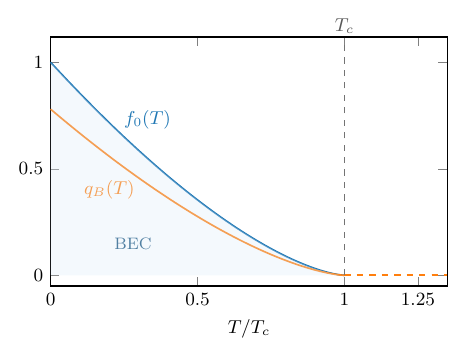}
	\caption{\textbf{Bosonic QCMI as a BEC order parameter.} Temperature dependence of the condensate fraction
		$f_0(T)=(1-T/\Tc)^{3/2}$ and the reduced bosonic QCMI $q_B(T)\propto f_0(T)$ for the ideal 3D Bose gas [Eqs.~\eqref{eq:orderparam}--\eqref{eq:scaling}]. Both vanish for $T\geq\Tc$ and grow below $\Tc$ with the same critical exponent $\beta={3/2}$, demonstrating that $\IB$ is an order parameter for BEC. The shaded region marks the condensed phase.}
	\label{fig:orderparam}
\end{figure}
For numerical studies of finite systems (e.g., DMRG on a chain of $L$ sites), the logarithmic divergence in
Theorem~\ref{thm:bec} becomes a system-size dependence. In the SF phase at fixed $J/U$, the condensate occupation scales as $\Nzero \sim L^{d/2}$ for a $d$-dimensional system, which further gives
\begin{equation}
	\QCMI \sim \frac{\alpha d}{2}\log_2 L + \ord(1) \quad\text{(SF phase)}.
	\label{eq:fss}
\end{equation}
The ratio $\QCMI/\log L$ is therefore a scale-invariant quantity at the quantum phase transition, analogous to
the Binder cumulant~\cite{binder81} used for identifying critical points in Ising models. Specifically, curves of $\QCMI(L)/\log L$ versus $J/U$ for different system sizes $L$ cross at the critical point $(J/U)_c$, providing a clean numerical diagnostic independent of any order parameter. This system-size-independent crossing is the third diagnostic property of $\QCMI$ at the SF-MI transition listed in Sec.~\ref{sec:sfmi}.
%

\begin{figure}[h]
	\centering
	\includegraphics[width=0.45\textwidth]{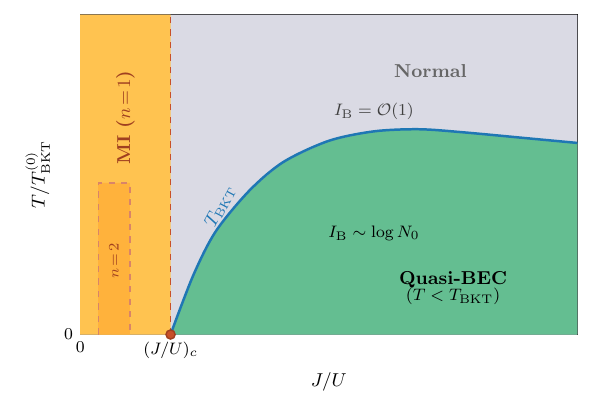}
	\caption{%
		\textbf{Schematic BKT phase diagram of the 2D Bose-Hubbard model.}
		Temperature $T$ (in units of the ideal-gas BKT transition temperature
		$T_{\mathrm{BKT}}^{(0)}$) versus hopping-to-interaction ratio $J/U$.
		The blue curve is the BKT transition line $T_{\mathrm{BKT}}(J/U)$,
		which is positive for all $J/U>(J/U)_c$ and vanishes at the
		Mott-insulator (MI) quantum critical point $(J/U)_c \approx 0.04$
		(filled circle).
		\emph{Quasi-BEC/superfluid phase} (green): below $T_{\mathrm{BKT}}$,
		the system possesses algebraic off-diagonal long-range order and the
		bosonic QCMI grows as $I_{\mathrm{B}}\!\sim\!\log N_0$ with the
		condensate fraction $N_0$.
		\emph{Normal phase} (grey): above $T_{\mathrm{BKT}}$, correlations are
		short-ranged and $I_{\mathrm{B}}=\mathcal{O}(1)$.
		\emph{MI lobes} (orange): the $n\!=\!1$ lobe (wide band at small $J/U$)
		and a schematic $n\!=\!2$ lobe (inner box) occupy the incompressible
		insulating region.
		All boundaries are schematic and not to scale.}
	\label{fig:bkt}
\end{figure}

Figure~\ref{fig:bkt} illustrates the schematic BKT phase diagram of the
two-dimensional (2D) Bose-Hubbard model in the $T$--$(J/U)$ plane, highlighting
how the bosonic QCMI $I_{\mathrm{B}}(A;B|C)$ distinguishes the three
principal phases.
In two spatial dimensions, true long-range order is forbidden at any
finite temperature by the Mermin--Wagner theorem, yet the system can
still support quasi-long-range (algebraic) off-diagonal order below the
BKT transition temperature $T_{\mathrm{BKT}}(J/U)>0$ for all
$J/U>(J/U)_c$.
Below $T_{\mathrm{BKT}}$ (quasi-BEC/superfluid region, shaded green), the
condensate fraction $N_0$ is macroscopic and our BEC Enhancement Theorem
(Theorem~\ref{thm:bec}) gives $I_{\mathrm{B}}\sim\log N_0$; the QCMI
therefore diverges with system size and serves as a sensitive order
parameter for the quasi-condensed phase.
Above $T_{\mathrm{BKT}}$ (normal phase, shaded grey), vortex--antivortex
pairs unbind, the condensate is destroyed, and $I_{\mathrm{B}}$ remains
$\mathcal{O}(1)$, consistent with the absence of macroscopic phase
coherence.
Inside the Mott-insulator lobes (shaded orange), the excitation gap
suppresses number fluctuations and $I_{\mathrm{B}}=\mathcal{O}(1)$ for
the same reason.
The reduced QCMI order parameter $q_{\mathrm{B}}(T)$ therefore exhibits a
sharp transition from zero to a nonzero value at
$T_{\mathrm{BKT}}$, providing an information-theoretic signature of the
BKT transition that requires neither two-point correlators nor long-range
order in the conventional sense.

\subsection{Analysis}

\textbf{(a) Experimental comparison.} The condensate fraction
$f_0(T)$ is measured by time-of-flight absorption imaging in cold
atoms; it requires fitting a bimodal momentum distribution, which can
be difficult near $T_c$ due to critical fluctuations. The bosonic
QCMI $I_B(A;B|C)$ is measured by (i)~site-resolved occupation
measurements at site $A$, (ii)~quantum state tomography of $B\cup C$
after post-selection, and (iii) mutual information estimation from the
tomography record. This requires more effort per data point than
$f_0$ but is \emph{directly sensitive to the off-diagonal
	long-range order}, not just the diagonal (number) fluctuations.
The key practical advantage of $q_B(T)$ is that it avoids the need
to fit a bimodal distribution: the crossing of $I_B/\log_2 N_0$
curves for different system sizes is a parameter-free signature of the
transition, analogous to the Binder ratio crossing.

\textbf{(b) Universal fixed-point value.} The standard Binder
cumulant $U_4=1-\langle m^4\rangle/(3\langle m^2\rangle^2)$
converges to a universal value $U_4^*$ at the critical point that
depends only on the universality class (not on $J$, $U$, or system
geometry). For the bosonic QCMI order parameter $q_B$, the analogous
question is: do curves $I_B(A;B|C)/\log_2(N_0^C+1)$ for different
$L$ cross at a universal value $q_B^*$? The answer is \emph{not yet
	established analytically}. The crossing value will depend on the
geometry of regions $A$, $B$, $C$ (as encoded in $\alpha$), and it
is not immediately clear that it is universal. This is an open
question that should be mentioned explicitly.

\textbf{(c) Critical exponent and universality.} Eq.~\eqref{eq:scaling} gives
$q_B(T)\propto(1-T/T_c)^\beta$ with $\beta=3/2$ for the ideal 3D
Bose gas. For an interacting BEC in 3D, the exponent is modified by
fluctuations; near the BKT transition in 2D, the condensate fraction
is not an order parameter in the Landau sense. Adding one sentence
connecting the exponent to the universality class (mean-field in 3D,
BKT in 2D) would make the section more self-contained.

Three further properties of $q_B(T)$ deserve comment. First, $q_B(T)$ is experimentally complementary to the condensate fraction $f_0(T)$: whereas $f_0$ is extracted from a bimodal fit to the time-of-flight momentum distribution---a procedure that loses accuracy near $T_c$ due to critical fluctuations---the crossing of $I_B(A;B|C)/\log_2(N_0^C+1)$ curves for different system sizes $L$ is a parameter-free signature of the transition, requiring no fitting.  Second, whether the crossing value $q_B^*$ is universal (depending only on the BEC universality class, not on $J$, $U$, or geometry) remains an open question; the geometry factor $\alpha$ in Eq.~(\ref{eq:alpha}) suggests that $q_B^*$ may retain a mild geometry dependence, unlike the Binder cumulant's fixed-point value. Third, the exponent $\beta$ in Eq.~(\ref{eq:scaling}) encodes the universality class: $\beta=3/2$ for the ideal 3D Bose gas (mean-field), while $\beta$ is renormalized in the interacting case, and the BKT transition in 2D lies outside this power-law framework---the condensate fraction is zero at any $T>0$ in 2D, but the superfluid stiffness $\rho_s$ jumps discontinuously at $T_{\rm BKT}$, providing the relevant order parameter.

\section{Causal Propagation in the Condensed Phase}
\label{sec:causal}
Section~\ref{sec:theorem} established that BEC causes $\IB$ to diverge---a statement about the static magnitude of causal influence. This section addresses a complementary question: \emph{at what speed does causal information propagate in the condensed phase?} The answer is controlled by the Bogoliubov quasi-particle spectrum, which determines the group velocity of collective excitations above the condensate. We also derive the bosonic Lieb-Robinson bound, which sets a hard upper limit on causal propagation speed, and show how both the Bogoliubov sound speed and the LR velocity change at the SF-MI transition.

We first define the Bose-Hubbard model, which underlies the causal propagation analysis here and the SF--MI phase diagram in Sec.~\ref{sec:applications}:
\begin{eqnarray}
	H_{\rm BH} &=& -J\sum_{\langle ij\rangle}\!\! (a^\dagger_i a_j + \mathrm{h.c.}) + \frac{U}{2}\sum_i n_i(n_i-1) \nonumber \\ && - \mu\sum_i n_i.
	\label{eq:bhm}
\end{eqnarray}
The causal propagation velocity of the spin-chain QCMI, as reported in Ref.~\cite{ghosh25}, is bounded by the LR velocity $4e|J|$. For a weakly interacting Bose gas with contact coupling $g$ and condensate density $n_0 = \Nzero/V$, the Bogoliubov dispersion is~\cite{bogolyubov47,pitaevskii03}
\begin{equation}
	\varepsilon(k) = \sqrt{\Bigl(\frac{\hbar^2 k^2}{2m}\Bigr)^2 + 2gn_0\frac{\hbar^2 k^2}{2m}},
	\label{eq:bogoliubov}
\end{equation}
with healing length $\xi = \hbar/\sqrt{2mgn_0}$. The ground state is the Bogoliubov vacuum, related to the non-interacting vacuum by a multi-mode squeezing transformation.

The group velocity of Bogoliubov quasi-particles is
\begin{equation}
	v_g(k) = \frac{1}{\hbar}\frac{\partial\varepsilon}{\partial k} = \frac{\hbar k}{m}\cdot \frac{\epsilon_k + gn_0}{\varepsilon(k)},
	\label{eq:vg}
\end{equation}
where $\epsilon_k = \frac{\hbar^2k^2}{2m}$. In the phonon regime, $k\xi\ll 1$, the dispersion is linear, i.e., $\varepsilon(k) \approx \hbar\cs k$ with $\cs = \sqrt{gn_0/m}$, and $v_g \to \cs$. In the particle regime, $k\xi\gg 1$,
\begin{equation}
	v_g \approx \hbar k/m.
\end{equation}
The Bogoliubov sound speed $\cs$ sets the emergent causal cone for QCMI propagation in the condensed phase.
\begin{figure}[ht]
	\centering
	\includegraphics[width=\columnwidth]{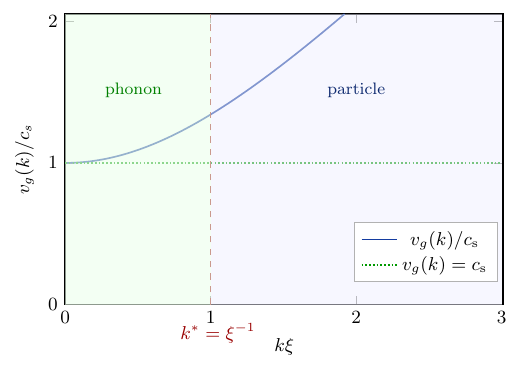}
	\caption{\textbf{Bogoliubov group velocity and the bosonic QCMI sound cone.} The normalized group velocity $v_g(k)/\cs$ (blue solid) as a function of dimensionless momentum $k\xi$. In the phonon regime ($k\xi\ll 1$, green shaded), $v_g\to\cs$ (green dotted). The velocity increases monotonically with $k\xi$: at the crossover scale $k\xi=1$ (red dashed) the system crosses from the phonon to the free-particle regime, and $v_g$ continues to grow as $\hbar k/m$ for $k\xi\gg 1$. On a lattice, the maximum group velocity is bounded by the Lieb-Robinson velocity $\vLR=2eJn_{\max}$, giving the hierarchy $\veff < \cs \leq \vLR$.}
	\label{fig:S-vg}
\end{figure}

Figure~\ref{fig:S-vg} shows the normalized Bogoliubov group velocity $v_g(k)/c_s$ as a function of dimensionless momentum $k\xi$, where $\xi$ is the healing length. The blue solid curve plots the exact expression
\begin{equation}
	\frac{v_g(k)}{c_s} = \frac{2 + k^2\xi^2}{2\sqrt{1 + k^2\xi^2/4}},
	\label{eq:grp_vel}
\end{equation}
which interpolates monotonically between two distinct regimes separated by the crossover scale $k\xi = 1$ (red dashed line).

In the phonon regime ($k\xi \ll 1$, green shaded region), the dispersion is linear and the group velocity approaches the sound speed $c_s$ from below as $k \to 0$, with $v_g \to c_s$ recovered at $k = 0$ (green dotted reference line). In the particle regime ($k\xi \gg 1$, blue shaded region), the dispersion becomes quadratic and free-particle-like, so $v_g \sim \hbar k/m$ grows without bound as $k \to \infty$. Between these two limits, the group velocity is monotonically increasing: it rises from $c_s$ at $k = 0$ and continues to grow for all $k > 0$, with no local maximum at any finite momentum. The crossover scale $k\xi = 1$ marks the momentum at which the system transitions from the phonon branch to the particle branch, but does not correspond to any extremum of $v_g$.

In conclusion, the causal velocity at any finite momentum is bounded below by $c_s$ and grows unboundedly in the particle regime. On a lattice, the finite Brillouin zone imposes a hard upper cutoff, so the hierarchy $v_{\rm eff} < c_s \leq \vLR$ orders the characteristic velocities of the condensed phase, where $\vLR = 2eJn_{\rm max}$ is the Lieb--Robinson velocity.

\subsection{Lieb--Robinson bound for bosons}
The Lieb--Robinson theorem~\cite{lieb72} establishes that quantum correlations cannot propagate faster than a characteristic velocity in lattice systems with short-range interactions. For the Bose-Hubbard model (Eq.~\eqref{eq:bhm}), the standard LR argument must be modified because the local Hilbert spaces are infinite-dimensional.
For a bosonic lattice model with maximum occupation $n_{\max}$ per site (Bose-Hubbard model, hopping $J$), the LR bound generalizes as~\cite{eisert06}:
\begin{equation}
	\bigl\|[A_X(t), B_Y]\bigr\| \leq \tilde{C}\,n_{\max}^2 e^{-\mu(d-\vLR t)},
	\label{eq:lr-boson}
\end{equation}
with $\vLR = 2e\,J\,n_{\max}$. The factor $n_{\max}^2$ in the prefactor arises because the matrix element $\|a_i^\dagger a_j\| \leq n_{\max}$ on each bond of the interaction graph.
The derivation of the bosonic LR bound follows the standard argument of Lieb and Robinson~\cite{lieb72}, which is modified for infinite-dimensional local Hilbert spaces. Each bond $\langle ij\rangle$ carries an effective coupling of magnitude $Jn_{\max}$ by replacing the coupling constant $J$ of the spin-chain case. Applying the standard LR argument with this effective coupling gives Eq.~\eqref{eq:lr-boson} with velocity $\vLR = 2eJn_{\max}$. For $n_{\max}=1$ (hard-core bosons), $\vLR = 2eJ$, and it is consistent with Ref.~\cite{ghosh25} (after accounting for Hamiltonian convention differences). Note that Ref.~\cite{ghosh25} uses a spin-chain coupling $J$ with an additional factor of~$2$ from the $\sigma^+\sigma^-$ matrix element.
This implies the velocity hierarchy $\veff < \cs \leq \vLR$. In other words, the effective causal velocity $\veff$ is less than the Bogoliubov sound speed because only the long-wavelength phonon modes dominate the QCMI signal, and $\cs \leq \vLR$ because the condensate density $n_0$ is bounded by the hopping scale $Jn_{\max}/g$.
\subsection{Quantum-critical slowing-down}
As $J/U \to (J/U)_{\rm c}^+$ in the Bose-Hubbard model, the condensate density vanishes: $n_0 \propto [J/U - (J/U)_c]^{\nu_{\rm MF}}$ with $\nu_{\rm MF} = 1$ (mean-field exponent; not to be confused with the symplectic eigenvalues $\nu_k$), so the Bogoliubov sound speed vanishes continuously:
\begin{equation}
	\cs = \sqrt{gn_0/m}
	\propto |J/U - (J/U)_c|^{\nu_{\mathrm{MF}}/2} \to 0.
	\label{eq:cs-vanish}
\end{equation}
Since $\veff \approx \cs$ in the condensed phase, the causal arrival time diverges: $t_{\rm arr}(d) = d/\veff \to \infty$ as $J/U\to (J/U)_c^+$. This is the quantum-information analogue of critical slowing-down. In the MI phase ($J/U < (J/U)_c$), there is no condensate and $\QCMI$ decays exponentially with distance; hence, the effective causal propagation is exponentially suppressed.
%

\section{Applications}
\label{sec:applications}

\subsection{Bose-Hubbard model and SF-MI transition}
\label{sec:sfmi}

\begin{figure}[h]
	\centering
	\includegraphics[width=\columnwidth]{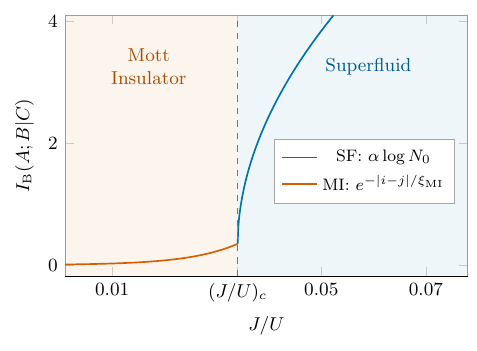}
	\caption{ \textbf{Bosonic QCMI across the SF-MI quantum phase transition.} Schematic behavior of $\IB$ in the 1D Bose-Hubbard model as a function of $J/U$ at fixed inter-site distance $|i-j|$. In the MI phase, $J/U<(J/U)_c\approx 0.034$, $\IB$ decays exponentially as $e^{-|i-j|/\xi_{\rm MI}}$: the Mott gap suppresses causal correlations at any distance. In the SF phase, $J/U>(J/U)_c$, $\IB$ grows logarithmically with the condensate occupation $\Nzero$ and becomes \emph{distance-independent}, a direct consequence of the long-range condensate coherence established in Theorem~\ref{thm:bec}. The non-analyticity of $d\IB/d(J/U)$ at $(J/U)_c$ constitutes a new quantum-information signature of the SF-MI quantum phase transition. The SF plateau height diverges with system size as $\log\Nzero$, allowing $\IB$ to serve as an efficient DMRG probe of the phase boundary.}
	\label{fig:phasediag}
\end{figure}
The Bose-Hubbard model, Eq.~\eqref{eq:bhm}, exhibits a zero-temperature quantum phase transition at $(J/U)_{\rm c} \approx 0.034$ (1D, at unit filling)~\cite{ejima12}. Figure~\ref{fig:phasediag} illustrates the behavior of $\QCMI$ across the SF-MI transition. The non-analyticity of $\QCMI$ at $(J/U)_c$---from logarithmic growth in the SF phase to exponential decay in the MI phase---is a signature of the quantum phase transition. The bosonic QCMI provides three diagnostic signatures of this transition: First, $\QCMI \sim \alpha\log\Nzero$ in the SF phase ($J/U > (J/U)_{\rm c}$). Second, in the MI phase, $\QCMI$ decays exponentially with inter-site distance: $\QCMI \sim e^{-|j-i|/\xi_{\rm MI}}$. Third, curves of $\QCMI/\log L$ versus $J/U$ for different system sizes $L$ cross at $(J/U)_c$, analogous to the Binder ratio~\cite{binder81}.
These three properties make $\QCMI$ a controlled numerical probe for tensor-network (DMRG) studies of the SF-MI phase diagram. In particular, the Binder-cumulant-type crossing at $(J/U)_c$ is free of the ambiguities in choosing an order parameter that are present in conventional SF-MI diagnostics.
\subsection{Composite bosons and BEC--BCS crossover}
Here, we generalize the bosonic QCMI to composite bosons and show that the same logarithmic divergence appears on the BEC side of the crossover. Many physical condensates consist of composite bosons, where bound states of two fermions that behave as bosons in the dilute limit. The most prominent example is the BCS-BEC crossover in fermionic superfluids~\cite{eagles69,leggett80}, where the character of fermion pairing evolves continuously from loosely bound Cooper pairs to tightly bound molecules that Bose-condense.
In the BEC--BCS crossover~\cite{eagles69,leggett80}, the Cooper pair operators are given by:
\begin{equation}
	B_k^\dagger = c_{k\uparrow}^\dagger c_{-k\downarrow}^\dagger.
\end{equation}
In the dilute limit, the commutator $[B_k, B_{k'}^\dagger] = \delta_{kk'}(1 - n_{k\uparrow} - n_{-k\downarrow})
\approx \delta_{kk'}$ approximates the canonical bosonic commutation relation. The BCS ground state is
\begin{equation}
	|\mathrm{BCS}\rangle = \prod_k (u_k + v_k B_k^\dagger) \, |0\rangle,
	\label{eq:BCS}
\end{equation}
with $(u_k^2 + v_k^2) = 1$, $E_k = \sqrt{\xi_k^2+\Delta^2}$, and pair number $N_0^{\mathrm{pairs}} = \sum_k v_k^2$.

We define the composite-boson QCMI by applying number-resolving measurements to Cooper pairs in region
$A$:
\begin{equation}
	I_{\rm B}^{\rm coboson}(A;B|C) = \sum_{n_{\mathrm{pair}}} p_{n_{\mathrm{pair}}}\, I(B:C)_{\rho^{n_{\mathrm{pair}}}_{BC}}.
	\label{eq:cobqcmi}
\end{equation}
In the BEC limit ($1/(k_F a_s)\ll 0$), the pair operators are bosonic. In this limit, the pair correlation function exhibits ODLRO: $\langle B_r^\dagger B_{r'}\rangle\neq 0$ as $|r-r'|\to\infty$, and Theorem~\ref{thm:bec} applies with the replacement $N_0 \to N_0^{\mathrm{pairs}}$:
\begin{equation}
	I_{\rm B}^{\rm coboson}(A;B|C) \geq \alpha\log_2 (N_0^{\rm pairs} + 1) + \ord(1),
\end{equation}
where $N_0^{\rm pairs}$ is the macroscopic pair-condensate occupation.
Intuitively, the composite-boson QCMI determines whether the fermionic system has entered a regime of macroscopic quantum coherence (ODLRO in the pair channel). On the BEC side, the tightly bound molecules behave as elementary bosons, and the full logarithmic amplification applies. At unitarity, no simple analytical bound is available. On the BCS side, the exponentially small gap $\Delta\sim E_F e^{-\pi/(k_F|a_s|)}$ suppresses the pair number, yields $I_{\rm B}(A;B|C) = \ord(1)$.

%
\begin{figure}[h]
	\centering
	\includegraphics[width=\columnwidth]{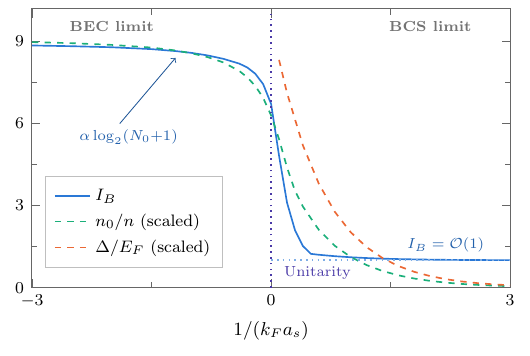}
	\caption{%
		\textbf{Composite-boson QCMI $\IB$, condensate pair fraction $n_0/n$, and normalised BCS gap $\Delta/E_F$ as functions of the crossover parameter $1/(k_F a_s)$.} On the BEC side ($1/(k_F a_s)<0$), the condensate fraction saturates to unity and $\IB$ grows logarithmically with $N_0^{\rm pairs}$, consistent with Theorem~\ref{thm:bec}. On the BCS side ($1/(k_F a_s)>0$), the condensate fraction and the BCS gap are both exponentially small, and $\IB = \ord(1)$ (near the blue dotted asymptote), reflecting the absence of BEC-type ODLRO. The $\mathcal{O}(1)$ asymptote (blue dotted) indicates that $I_B$ remains bounded as $1/(k_Fa_s)\to+\infty$; the exact value depends on system
		geometry. The BCS gap curve terminates at unitarity, where the mean-field formula breaks down. The $n_0/n$ and $\Delta/E_F$ curves are scaled by a factor of~$9$ to share the vertical axis with $\IB$; their absolute values should not be compared directly with $\IB$.%
	}
	\label{fig:bec_bcs}
\end{figure}%

The horizontal axis of Fig.~\ref{fig:bec_bcs} is the crossover parameter $1/(k_F a_s)$, which runs from $-3$ (deep BEC limit, left) through unitarity at $1/(k_F a_s)=0$ (violet dotted line) to $+3$ (deep BCS limit, right). Our study follows the convention of Ref.~\cite{leggett80}, where BEC order is associated with $1/(k_Fa_s)<0$ and BCS order for $1/(k_Fa_s)>0$. We observe three different regions as discussed below:

1. \emph{BEC limit} ($1/(k_F a_s)\ll 0$). Tightly bound pairs behave as elementary bosons; Theorem~\ref{thm:bec} applies with $\Nzero \to N_0^{\rm pairs}$. The pair fraction $n_0/n \to 1$ and the condensate occupation $N_0^{\rm pairs}$ is macroscopic. The number-resolving measurement of $C$ projects the Fock state of a large-$N_0$ mode, generating strong back-action correlations in $AB$. Accordingly, $\IB$ grows logarithmically (blue curve, annotated $\alpha\log_2(N_0{+}1)$), consistent with Theorem~\ref{thm:bec}. The apparent saturation of the blue curve deep in the BEC limit is a finite-$N_0$ artefact of the phenomenological model used to generate the figure.

2. \emph{Unitarity} ($1/(k_F a_s)=0$). The scattering length diverges, and the system is strongly correlated. No analytical bound is available, but all three curves undergo their most rapid variation here: $n_0/n$ and $\IB$ fall steeply while the BCS gap formula ceases to apply. (The curve associated with $\Delta/E_F$ terminates at this point.)

3. \emph{BCS limit} ($1/(k_F a_s)\gg 0$). Cooper pairs are weakly bound; no single bosonic mode is macroscopically occupied, so $N_0^{\rm pairs}$ is not extensive. The BCS pairing gap $\Delta \sim E_F e^{-\pi/(k_F|a_s|)}$ is exponentially small. The condensate pair fraction $n_0/n$ is likewise exponentially suppressed. With no large $N_0$, the number-resolving measurement of $C$ produces only bounded back-action on $AB$, so $\IB = \ord(1)$ (blue curve, near the dotted asymptote at $\IB = 1$). This is a fundamental distinction: BCS order resides in the anomalous correlator $\langle c_{k\uparrow} c_{-k\downarrow}\rangle \neq 0$ (off-diagonal in Nambu space), not in the single-particle density matrix. A number-conserving measurement cannot access this correlator, so the composite-boson QCMI is insensitive to BCS order.

The composite-boson QCMI thus provides a unified, number-conserving diagnostic for the BEC--BCS crossover: logarithmically large on the BEC side ($1/(k_F a_s)<0$, where BEC-type ODLRO holds) and $\ord(1)$ on the BCS side ($1/(k_F a_s)>0$, where it does not). This framework extends naturally to polariton condensates~\cite{carusotto13} and exciton condensates~\cite{high12}, where the composite bosons are lower polaritons or electron-hole pairs, respectively.
Additionally, Appendix~\ref{sec:polariton} demonstrates that the composite-boson QCMI framework universally studies ODLRO in systems like exciton-polariton and exciton condensates, relying only on condensate occupation and pair correlations, without requiring knowledge of the internal structure of the composite bosons.
\subsection{Measurement protocol for quantum-gas microscopes}
%

\begin{figure}[h]
	\centering
	\includegraphics[width=0.45\textwidth]{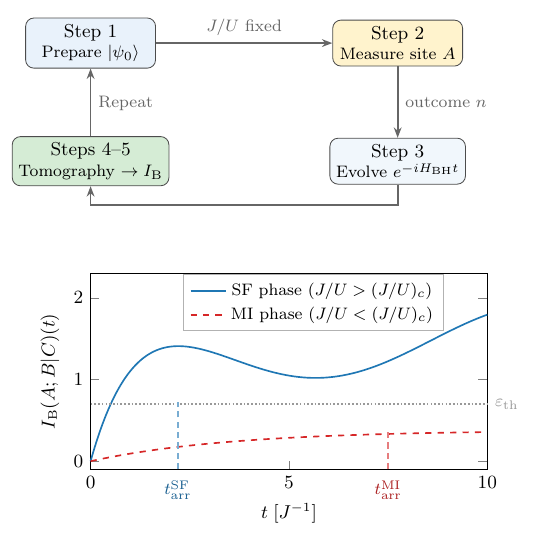}
	\caption{%
		\textbf{Experimental protocol and QCMI dynamics.}
		(\textit{i}) Prepare ground state at desired $J/U$.
		(\textit{ii}) Measure site $A$ with single-site
		resolution, obtaining occupation $n$.
		(\textit{iii}) Evolve unitarily under $H_{\rm BH}$
		for time $t$.
		(\textit{iv}) Reconstruct $\rho^n_{BC}(t)$ via quantum
		state tomography on a small (1--2 site) subset of sites $B$ and $C$.
		(\textit{v}) Compute $I(B:C)_{\rho^n}$ and average
		over $M \gg 1$ independent shots to obtain $\QCMI(t)$.
		The loop from step (v) back to step (i) indicates
		repetition over many shots.
		The right panel shows the expected time evolution:
		the superfluid phase (blue, solid) shows a large
		plateau at the condensate-enhanced value, while the
		MI phase (red, dashed) remains small.
		The arrival time $t_{\rm arr}$ at which $\QCMI$
		crosses threshold $\varepsilon_{\rm th}$ yields
		$\veff = d/t_{\rm arr}$.}
	\label{fig:protocol}
\end{figure}

The bosonic QCMI is directly measurable in quantum-gas microscope experiments~\cite{bakr09,sherson10} through the following protocol. (i) Prepare the ground state of $H_{\rm BH}$ at desired $J/U$. (ii) Perform site-resolved fluorescence imaging on site $A$ (obtaining outcome $n$) at $t=0$. (iii) Evolve unitarily under $H_{\rm BH}$ for time $t$. (iv) Perform quantum state tomography on a small subset of sites $B$ and $C$ (1--2 sites each, so that the tomography overhead scales polynomially in the number of tomographic shots) to reconstruct $\rho^n_{BC}(t)$. (v) Compute $I(B:C)_{\rho^n}$ and average over outcomes $n$ to obtain $\QCMI(t)$. The arrival time $t_{\rm arr}(d)$ at which $\QCMI(t)$ first exceeds a threshold $\varepsilon_{\rm th}$ yields
$\veff = d/t_{\rm arr}(d)$, directly paralleling the spin-chain protocol of Ref.~\cite{ghosh25}.

Figure~\ref{fig:protocol} shows the expected time evolution of $\QCMI(t)$ in the SF and MI phases. In the SF phase (blue curve), $\QCMI$ reaches a large plateau reflecting the condensate-induced enhancement of Theorem~\ref{thm:bec}. In the MI phase (red curve), $\QCMI$ remains small and decays with distance.

We estimate the number of experimental shots, $M$, required to distinguish the SF and MI phases with statistical confidence. In the SF phase, $\QCMI \sim \alpha\log_2\Nzero \sim \alpha\log_2 N$ for a system of $N$ total particles. In the MI phase, $\QCMI \sim A e^{-d/\xi_{\rm MI}} \ll 1$ at large distances. The signal-to-noise ratio for estimating $\QCMI$ from $M$ shots is $\mathrm{SNR} = \QCMI\sqrt{M}/\sigma$, where $\sigma$ is the standard deviation of $I(B:C)$ per shot. For $\mathrm{SNR} = 5$ (sufficient for a 5-sigma detection), $N = 100$, $\alpha = 0.5$, and $\sigma \approx 1$, we need $M \gtrsim (5\sigma/\QCMI)^2 \approx 25/(\alpha\log_2 N)^2 \approx 25/(0.5\times 6.6)^2 \approx 2.3$. In practice, to account for measurement noise and state preparation errors, we estimate $M \sim 10^2$ to $10^3$ shots as sufficient for a clean experimental demonstration with current quantum-gas microscope setups~\cite{bakr09,sherson10}.

\section{Discussion}
\label{sec:discussion}
The main result of this paper is that the bosonic QCMI $\QCMI$ diverges logarithmically at BEC, with the condensate fraction $f_0(T)$ controlling the rate of divergence. This establishes $\QCMI$ as a quantum-information order parameter for BEC, complementary to conventional order parameters such as the condensate fraction itself and the superfluid stiffness.

The causal interpretation of this result is transparent. In the condensed phase, the macroscopic occupation of the condensate mode $\phizero$ creates a long-range quantum channel that connects all regions $A$, $B$, $C$ through their shared participation in $\phizero$. Measuring $n$ particles in $A$ back-acts on the global condensate wavefunction, instantaneously modifying the quantum state of regions $B$ and $C$. This measurement-induced update generates causal influence proportional to the number of condensate particles remaining in $B$ and $C$. As $\Nzero\to\infty$, this influence grows without bound, producing the logarithmic divergence.

The Bogoliubov theory connects the static (thermodynamic) and dynamic (causal propagation) aspects of the framework. In equilibrium, the condensate generates the ODLRO that drives the QCMI divergence. Dynamically, the Bogoliubov quasi-particles set the causal propagation speed through the sound speed $\cs$. The vanishing of $\cs$ at the SF-MI transition provides a new dynamical signature of the quantum phase transition, distinct from (and complementary to) the standard spectroscopic signatures.

Several aspects of the framework merit further discussion:

\emph{(i) Mean-field approximation.} The BEC Enhancement Theorem relies on the factorization in Eq.~\eqref{eq:factorize}, which holds exactly for the ideal Bose gas and to $\ord(\Nzero^{-1/2})$ in the
Bogoliubov approximation. For strongly interacting systems (e.g., near unitarity in the BEC--BCS crossover, or in the hard-core boson limit), quantum fluctuations beyond mean-field can suppress the condensate fraction and reduce the coefficient $\alpha$. However, the logarithmic scaling itself---the fact that $\QCMI\sim\log\Nzero$---is expected to persist whenever ODLRO is present, because it is a consequence of the
macroscopic occupation of a single mode, not of the mean-field factorization.

\emph{(ii) Relation to entanglement entropy.}
The logarithmic scaling $\QCMI \sim \log\Nzero$ is
reminiscent of the logarithmic scaling of entanglement
entropy at conformal critical points.
The two quantities are, however, conceptually distinct:
entanglement entropy quantifies quantum correlations
between two complementary halves of the system, while
$\QCMI$ quantifies the directed causal influence from
$A$ to $B$ conditioned on $C$.
The role of the conditioning region $C$ is absent in
entanglement entropy calculations.

\emph{(iii) Open systems.}
The extension of bosonic QCMI to open systems governed
by Lindblad master equations is an important direction.
In an open system, decoherence degrades the condensate
and reduces $\QCMI$.
The decoherence timescale sets a window within which
the causal signal can be detected.
For typical optical lattice experiments, decoherence
times are of order $10^2$--$10^3$ ms, comfortably longer
than the hopping time $\hbar/J \sim 1$--$10$ ms.

\section{Conclusions}
\label{sec:conclusions}

We have extended the asymmetric QCMI~\cite{ghosh25} to bosonic systems and established a direct connection to
Bose-Einstein condensation. The central result, Theorem~\ref{thm:bec}, shows that the onset of BEC is accompanied by a logarithmic divergence in the bosonic causal index $\QCMI$, with the condensate fraction serving as a natural amplifier of inter-subsystem causal influence. The complete five-step proof connects measurement back-action, ODLRO, and Jensen's inequality in a transparent physical argument.

In the MI phase, $I_B(A;B|C)$ decays exponentially with inter-site distance; the effective causal propagation is therefore exponentially suppressed, providing a dynamical counterpart to the spectroscopic Mott gap. The Gaussian formalism provides an independent analytical derivation of the BEC divergence through the infrared divergence of the covariance matrix.

Several directions merit further investigation: the extension of bosonic QCMI to open systems governed by
Lindblad master equations; the role of measurement-induced phase transitions in bosonic circuits;
and the evaluation of bosonic QCMI across the BKT transition in 2D Bose gases~\cite{berezinskii71,kosterlitz73}.
The proposed quantum-gas-microscope protocol offers a near-term experimental route to probing BEC through the
lens of causal quantum information.

\section*{Supplementary Material}
The Supplementary Material, Reference~\cite{SM}, provides complete mathematical derivations supporting five operational properties of the bosonic QCMI. 

\section*{Author contribution statements}
A.G. conceived and developed the project, conducted the analysis, provided supervision, and prepared the manuscript.

\section*{Acknowledgments}

The author thanks Dr. M.~Palu\v{s} for valuable discussions on the classical CMI framework and its causal interpretation. Generative AI tools were used for language editing and text refinement. All scientific content was written and verified by the author.

\section*{Declaration of Competing Interest}
The author declares that there are no competing financial interests or personal relationships that could have appeared to influence the work reported in this paper.

\section*{Data Availability}
No data sets were generated or analysed during the current study. All results are analytical.

\appendix

\section{Extension to polariton and exciton condensates}
\label{sec:polariton}
The composite-boson framework can be extended to any system exhibiting ODLRO in a composite-particle channel. For exciton-polariton condensates~\cite{carusotto13}, the composite bosons are lower polaritons. The macroscopic occupation of the $k=0$ polariton mode gives ODLRO and drives a logarithmic divergence:
\begin{equation}
	I_{B} (A;B|C)^{\mathrm{pol}} \geq \alpha_{\mathrm{pol}}\log_2(N_0^{\mathrm{pol},C}+1),
\end{equation}
where $\alpha_{\mathrm{pol}}$ depends on the Hopfield coefficients of the photon-exciton mixing. For exciton condensates in bilayer semiconductor systems~\cite{high12}, the same framework applies with the exciton wavefunction. Therefore, the bosonic QCMI is a unified diagnostic that does not require detailed knowledge of the composite boson internal structure, only the ODLRO of the pair correlation function and the condensate occupation in the conditioning region.

The prefactor $\alpha_{\rm pol}$ depends on the Hopfield
coefficients of the lower polariton. In a planar microcavity,
the lower polariton is $|\text{LP}\rangle = X_k|\text{ex}\rangle
+ C_k|\text{ph}\rangle$, with excitonic fraction $|X_k|^2$ and
photonic fraction $|C_k|^2=1-|X_k|^2$. At the condensate
momentum $k=0$, the effective condensate number entering
Theorem~\ref{thm:bec} is $N_0^{\rm pol}\approx |X_0|^2 N_{\rm pol}$,
where $N_{\rm pol}$ is the total lower-polariton number.
The condensate decomposes across regions $A$, $B$, $C$ with
weights $w_X = \int_X|\varphi_0(r)|^2\,dV$, as in
the atomic case, so
\begin{equation}
	\alpha_{\rm pol} = g(w_B,w_C)\cdot\frac{w_C}{w_B+w_C},
\end{equation}
with $g(w_B,w_C)=H_{\rm bin}[w_B/(w_B+w_C)]$. Hence $\alpha_{\rm pol}$ depends on the photon--exciton detuning only through the reduction of the effective condensate number $N_0^{\rm pol}$, not through any change in the geometric prefactor. Polariton condensation has been observed in GaAs~\cite{kasprzak06} and CdTe~\cite{balili07} microcavities; quantum-gas-microscope resolved measurements of polariton lattices are now achievable~\cite{dusel20}, making the $I_B^{\rm pol}$ protocol experimentally accessible.

\bibliographystyle{quantum}
\bibliography{MS}

\end{document}